\documentclass[twocolumn,aps,prl,amsmath,amssymbpreprintnumbers]{revtex4-1}
\usepackage[normalem]{ulem}
\usepackage{hyperref}
\begin{document}

\title{Loops rescue the no-boundary proposal}

\author{Martin Bojowald} 
\email{bojowald@gravity.psu.edu}
\affiliation{Institute for Gravitation and the Cosmos, The Pennsylvania State
  University, 104  Davey Lab, University Park, PA 16802, USA}

\author{Suddhasattwa Brahma}
\email{suddhasattwa.brahma@gmail.com}
\affiliation{Asia Pacific Center for Theoretical Physics, Pohang 37673, South Korea}

\begin{abstract}
  New and non-trivial properties of off-shell instantons in loop quantum
  cosmology show that dynamical signature change naturally cures recently
  observed problems in the semiclassical path integral of quantum gravity. If
  left unsolved, these problems would doom any theory of smooth initial
  conditions of the universe. The no-boundary proposal, as a specific example
  of such a theory, is rescued by loops, presenting a rare instance of a
  fruitful confluence of different approaches to quantum cosmology.
\end{abstract}

\maketitle

A major ambition of quantum cosmology is to derive properties of the universe
from a theory of quantum gravity combined with a specific selection of initial
conditions at the big bang. There have been two main proposals of initial
conditions, the no-boundary wave function \cite{nobound} and the tunneling
picture \cite{tunneling}. They have been unified recently, but also came under
new scrutiny \cite{LorentzianQC,RealNoBound} by an application of
Picard--Lefschetz theory to the Lorentzian path-integral of quantum
gravity. An extension to perturbative and non-perturbative deviations from
spatially isotropic geometries has revealed two possible outcomes, each of
which would eliminate the attraction of a theoretical proposal setting
conditions at the very beginning of the universe: Either there are run-away
perturbations \cite{NoSmooth,NoRescue,NewNoBound,NoBoundReg}, or, at the very
least, the initial conditions have to be amended by hand so as to achieve
stability, a late-time property \cite{DampedNoBound}. The problem is very
general and can be traced back to geometrical properties of space-time in
general relativity. Only a modified space-time structure can overcome such an
obstruction. As we will see, this happens rather naturally in loop quantum
cosmology.

A different scenario has been developed for some time in loop quantum
cosmology \cite{LivRev,ROPP}. Initially, in exactly homogeneous cosmological
models there were several indications \cite{Sing,GenericBounce,QuantumBigBang}
that quantum space-time effects could lead to a bounce at large (Planckian)
density, such that the expanding part of the universe may have been preceded
by collapse. The inclusion of inhomogeneity in a covariant fashion, as
perturbations \cite{ScalarGaugeInv,ScalarHolInv} or non-perturbatively within
spherical symmetry \cite{JR,HigherSpatial,WDWSS}, then revealed an unforeseen
implication of the same space-time effects that resolve the singularity: At
large density, the universe has the structure of a certain 4-dimensional
Euclidean geometry in which the usual time direction of Lorentzian space-time
is replaced by a fourth spatial dimension \cite{Action,SigChange}. Crucially,
this dynamical signature change, unlike classical versions, is non-singular
\cite{SigImpl}.

Nothing can propagate if there is only space, and therefore the original
bounce picture is altered. However, the emergence of Euclidean space is
strikingly similar to the technical implementation of the no-boundary
proposal, which posits that the classical form of expanding space-time is
``rounded off'' by a Euclidean cap replacing the big-bang singularity.
In this letter, we show that the analogy is not just a formal one: The main
difficulties of the no-boundary proposal, found in \cite{NoSmooth,NoRescue},
can be resolved if the Lorentzian path integral is augmented by some of the
quantum-geometry effects found in loop quantum cosmology. In particular,
off-shell instantons in loop quantum cosmology are subject to signature change
{\em even if the energy density is significantly sub-Planckian.}

The specific forms in which Euclidean space makes its appearance in the
no-boundary proposal and in loop quantum cosmology, respectively, are quite
different from each other. In the no-boundary proposal, the origin of
Euclidean space is a combination of the formal Wick rotation often employed in
quantum-field theory --- replacing real time $t$ with complex time
$\tilde{t}=\pm it$ --- with the selection of specific saddle points to
evaluate a semiclassical path integral. As a consequence of Wick rotation, a
standard cosmological line element becomes Euclidean in the new coordinate
$\tilde{t}$, as in
\begin{eqnarray} \label{ds}
 {\rm d}s^2&=&-N(t)^2{\rm d}t^2 +a(t)^2{\rm d}\Omega_k^2
 = \tilde{N}(\tilde{t})^2{\rm d}\tilde{t}^2 +\tilde{a}(\tilde{t})^2{\rm
   d}\Omega_k^2
\end{eqnarray}
for an isotropic cosmological model with spatial line element ${\rm
  d}\Omega_k^2$, $k=0$ or $k=\pm 1$ indicating the curvature of space.  The
new scale factor, $\tilde{a}(\tilde{t})$, describes how constant-$\tilde{t}$
hypersurfaces grow as $\tilde{t}$ increases, but since $\tilde{t}$ has lost
its meaning as time, it no longer describes dynamical growth. The lapse
function $N(t)$ specifies the rate of progress of time $t$.  The way it
appears in (\ref{ds}) shows that a coordinate-independent Wick rotation can be
formulated by replacing $N$ with $\tilde{N}=\pm iN$, leaving $t$ unchanged.

{\em Loop quantization:} Loop quantum cosmology does not introduce complex
coordinates or complex lapse functions. Instead, it is based on quantum
modifications of spatial or space-time geometry which in the broader setting
of loop quantum gravity \cite{ThomasRev,Rov} have been found to facilitate the
technical implementation of a quantum theory of gravity in terms of a Hilbert
space and well-defined operators. In particular, the basic continuum
quantities of spatial geometry, such as areas and volumes, are represented by
operators with discrete spectra \cite{AreaVol,Area,Vol}. Moreover, an
infinitesimal change of these quantities in time --- or, more geometrically,
the extrinsic curvature of space in space-time --- no longer has a linear and
local expression in space but is instead exponentiated and extended
one-dimensionally, along an eponymous loop \cite{LoopRep,ALMMT}. In a
cosmological model such as (\ref{ds}), two new effects are implied
\cite{IsoCosmo}: (i) There is no operator that directly represents extrinsic
curvature $\dot{a}$ or the Hubble parameter $\dot{a}/a$, and (ii) an operator
version of $a^2$ has a discrete spectrum.

Effect (i) leads to holonomy modifications of cosmological equations: While
spatial non-locality is not visible in homogeneous cosmological models, the
specific form of non-linearity realized in loops implies that there are
operators only for periodic functions such as $\sin(\ell(a)\dot{a})/\ell(a)$
with a function (or constant) $\ell(a)$ depending on quantization
ambiguities. Any polynomial appearance of $\dot{a}$, as in the Friedmann
equation, should therefore be approximated by trigonometric holonomy functions
before it can be loop quantized. If $\ell(a)\approx \ell_{\rm P}/a$ with the
Planck length $\ell_{\rm P}$ (a choice made in \cite{APSII}), the new version
differs from the classical Friedmann equation only near Planckian
curvature. There are noticeable on-shell effects only around the big bang, but
there they may be significant.

A discrete spectrum as in effect (ii) leads to strong deviations from
continuity only for quantities sensitive to the spacing, which for an area
such as $a^2$ would be tiny (Planckian). However, another consequence is that
an operator with a discrete spectrum containing the zero eigenvalue, as is the
case for $a^2$, does not have an inverse operator, usually defined by
inverting the eigenvalues. The need to quantize $1/a^n$ for different choices of
$n$, for instance to obtain well-defined matter Hamiltonians, can be fulfilled
by using commutators \cite{QSDV} such as
\begin{equation} \label{Inverse}
 \hat{h}^{-1}[\hat{h},\sqrt{\hat{a}}]= -\frac{1}{2}\hbar \ell
\widehat{a^{-1/2}}
\end{equation}
where $\hat{h}=\widehat{\exp(i\ell \hat{p}_a)}$, using the momentum
$p_a\propto \dot{a}$, is a ``holonomy'' operator or a quantization of
$\dot{a}$ that exists in loop quantum cosmology. The left-hand side of
(\ref{Inverse}) does not require the non-existing inverse of an operator with
discrete spectrum containing zero, and yet an operator with an inverse power
of $a$ in the classical limit results on the right \cite{InvScale}. Using such
commutator versions implies stronger deviations from the continuum than the
spacing of the $a$-spectrum alone would indicate, in particular for small
values of $a$ which may be relevant near the big bang. An analysis of these
new operators \cite{Ambig,ICGC} shows that the eigenvalues $(a^{-1})_j$ of
well-defined inverse operators of $a$ are related to the eigenvalues $a_j$ of
$a$ by $(a^{-1})_j = f(a_j)/a_j$ with a function $f(a_j)$ that approaches
$f(a_j)=1$ for large $a_j$ and, for small $a_j$, is approximated by a
power-law form $f(a_j)\approx a_j^n$ with a positive integer $n>2$. The
small-$a$ behavior eliminates the divergence of a direct inverse at
$a_j=0$. The precise form of $f(a_j)$ and the small-$a$ power $n$ depends on
quantization ambiguities.

Both loop effects result from geometrical considerations independent of common
quantum effects such as fluctuations. In a first approximation, they can
therefore be studied with a modified version of the Friedmann equation (or the
corresponding Hamiltonian) in which any appearance of $\dot{a}$ is written
trigonometrically, and any inverse of $a$ is replaced by an appropriate power
of $a^{-1}f(a)$. In order to capture the main effects contained in the two
functions $\ell(a)$ and $f(a)$, we replace the classical Hamiltonian
(constraint) underlying the Friedmann equation,
\begin{equation} \label{C}
 C_{\rm class}=-\frac{3}{8\pi G} a\left(\frac{\dot{a}^2}{N^2}+k\right)+ m(a)=0
\end{equation}
using a generic matter energy $m(a)$, with
\begin{equation} \label{Cmod}
 C=-\frac{3}{2} \left(Q \frac{\sin^2(\delta
     P)}{\delta^2}+\frac{Q^{1/3}}{(4\pi G)^{-2/3}} \kappa(Q)\right)+ m(Q)g(Q)=0\,.
\end{equation}
Here, $Q=a^3/(4\pi G)$ and $P=-\dot{a}/(Na)$ are canonical variables.  We have
assumed a specific $a$-dependence of the first ambiguity function,
$\ell(a)=\delta a^{-1}$, $\delta \sim \ell_{\rm P}$, motivated by previous
studies that suggested a preference for this behavior; see, for instance,
\cite{ScalarHolInv}. Our results can be generalized to a power-law behavior
$\ell(a)\propto a^{2x}$, but independently indicate the same preference for
$x=-1/2$.

Suitable powers of the inverse-$a$ correction function $f(a)$ lead to a
function $\kappa(Q)$ in the curvature term ($\kappa(Q)=k$ classically), and to
a factor $g(Q)$ multiplying the matter energy $m(Q)$. Following the canonical
procedure, one then derives the modified Friedmann equation
\begin{eqnarray} \label{Friedmann}
 \left(\frac{\dot{a}}{Na}\right)^2 &=&  \left(\frac{8\pi G}{3}
   \frac{m(a)g(a)}{a^3}- 
   \frac{\kappa(a)}{a^2}\right)\\
&&\times \left(1+\delta^2
   \frac{\kappa(a)}{a^2}- \frac{m(a)g(a)}{a^3 \rho_{\rm QG}}\right) \nonumber
\end{eqnarray}
with $\rho_{\rm QG} = 3/(8\pi G \delta^2)$.

{\em Covariance:} 
Any modification of general relativity must respect covariance in the sense
that the degrees of freedom in the metric that correspond to coordinate
choices do not contribute to observable quantities. The Friedmann equation of
isotropic models tells us how the scale factor and matter change with respect
to a time coordinate $t$ that is compatible with the line element
(\ref{ds}). Coordinate time $t$ enters (\ref{C}) or (\ref{Friedmann}) only in
the form $N{\rm d}t$, a term which is invariant with respect to coordinate
changes $t\mapsto t'(t)$. Even after modifications by loop effects, such as
the $\delta$-term in (\ref{Friedmann}), the isotropic model remains time
reparameterization invariant.  But the equation shows no information about
covariance with respect to coordinate changes that mix space and time, such as
$t\mapsto t'(t,x_j)$ where $x_j$, $j=1,2,3$, are the spatial coordinates,
because such transformations do not preserve the form (\ref{ds}) of the line
element from which (\ref{C}) has been derived via Einstein's equation.

If one would like to test whether a modified Friedmann equation can be a part
of a generally covariant theory, one should at least include perturbative
inhomogeneity, such as the scalar and tensor modes prominent in
early-unniverse cosmology, coupled to the scale factor of an isotropic
background. In this setting, one can perform small coordinate changes of the
form $t\mapsto t'(t,x_j)$ in addition to $x_i\mapsto x_i(t,x_j)$.  Covariance
under these transformations turns out to be highly restrictive. There are now
four independent transformations, which must form a group or, in infinitesimal
form, an algebra. If we consider a sufficiently small region such that
space-time is locally Minkowski, the classical transformations are given by
the Poincar\'e algebra. Without perturbative inhomogeneity, by contrast, we
only have the much simpler algebra given by time translations.

Modifying the Friedmann equation implies a modification of time
translations. The issue of covariance, in a small, locally Minkowski region,
is whether the modified time translation can be a part of a consistent (and
perhaps modified) version of the Poincar\'e algebra. Canonical gravity
provides methods to test this condition for a modification such as
(\ref{Cmod}), which has been worked out in several models of loop quantum
gravity.

The result is that covariant perturbations are possible, but they are such
that the classical Poincar\'e algebra can no longer be used. In particular, at
large curvature (large momentum $\delta P$) there is a crucial sign change in
the commutator of a time translation with a boost. The same sign change is
obtained if we use the classical Poincar\'e algebra but replace the time
translation with a space translation, and the boost with a rotation. In this
way, the modification in (\ref{Cmod}) or (\ref{Friedmann}) for $\delta\not=0$
implies signature change when $\delta P$ is large.  In some cases, it can be
shown \cite{Normal,EffLine} that the redefined line element
\begin{equation} \label{dsSig}
 {\rm d}s_{\beta}^2 = -\beta N^2{\rm d}t^2+ a(t)^2{\rm d}\Omega_k
\end{equation}
instead of (\ref{ds}) is consistent with the modified Poincar\'e relations
where $\beta$ is the function that determines the sign change in the
time-boost commutator. When $\beta<0$ at large curvature, signature change is
explicit.

{\em Signature change:}
Signature change as it appears in models of loop quantum gravity is,
therefore, rather different from the version used in the no-boundary
proposal. We will now show that it can have important consequences in the same
context. In particular, as the main result, we can use a Lorentzian path
integral --- that is, an integral over paths weighted by $\exp(iS/\hbar)$
rather than $\exp(-S/\hbar)$ --- and have stable perturbations. To do so, we
need, as the major new ingredient, off-shell instantons which do not solve
(\ref{Friedmann}) but rather the second-order equation of motion of the
background theory. (No-boundary initial conditions are formulated with respect
to gauge-fixed time, such that the first-order equation is no longer
enforced. In this way, the Euclidean cap --- a non-classical property --- can
be realized in a semiclassical path integral.)  Such solutions then have to be
extended by a covariant theory of perturbative inhomogeneity.

In order to facilitate a comparison with derivations in \cite{NoRescue}, we
will assume that $8\pi G m/a^3=\Lambda$ is a cosmological constant, choose a
lapse function $N=M/a$ with constant $M$, write the modified Friedmann
equation in terms of $q=a^2$, and derive a second-order equation for this
variable. We will then solve the second-order equation, (\ref{qdd}) below,
without imposing the first-order one, (\ref{Friedmann}), thus dealing with
off-shell Lorentzian instantons, as is required to carry out the no-boundary
path integral. Several unexpected cancellations will make sure that the
function $\beta$ in (\ref{dsSig}) remains meaningful for off-shell
instantons. Moreover, and also surprisingly, such off-shell $\beta$ do not
require large densities to be negative. These new features are the reason why
loop quantum cosmology is able to rescue the no-boundary proposal.

Deriving the second-order equation for $q$ leads to
\begin{equation} \label{qdd}
 \ddot{q}=\frac{2}{3}\Lambda M^2 \left(1-\frac{3}{\Lambda} \frac{{\rm
       d}\kappa}{{\rm d}q}
   +\delta^2\left(1-\frac{3}{\Lambda}\frac{\kappa}{q}\right) \left(2\frac{{\rm
       d}\kappa}{{\rm d}q}-\frac{\kappa}{q}-\frac{\Lambda}{3}\right)\right)
\end{equation}
The right-hand side of (\ref{qdd}) is always regular in $q$ thanks to
inverse-$a$ corrections; the no-boundary initial value $q(0)=0$ can therefore
be imposed. For small $t$, $q$ and $\kappa/q$ are small and the right-hand
side of (\ref{qdd}) is approximately constant. A generic small-$t$ behavior of
$q(t)\propto t+O(t^2)$, as in \cite{NoRescue} but possibly with modified
coefficients, then follows.  A discussion of stability requires only the
small-$t$ behavior.

Identifying small $t$ with the early universe in models of loop quantum
gravity, one could expect that signature change is automatically
realized. However, this result is far from clear. Signature change has been
derived for on-shell solutions which obey the first-order Friedmann
equation. The second-order equation used for off-shell instantons has a larger
solution space to which standard result do not necessarily apply. Moreover,
signature change in models of loop quantum gravity requires strong quantum
space-time effects, which are usually found only at large, near-Planckian
curvature: The canonical analysis that ensures covariance implies that the
function $\beta$ to be used in (\ref{dsSig}) has the form
$\beta(P)=\cos(2\delta P)$ \cite{JR,ScalarHolInv,DeformedCosmo}. It takes the
value $\beta(P)=-1$ if $\delta P=\pi/2$, such that $\sin(\delta P)=1$ and
(\ref{Cmod}) implies Planckian matter density if $\delta\sim\ell_{\rm P}$.
One of the advantages of the no-boundary proposal, however, is that it might
be able to explain the origin of the universe without referring to uncertain
Planckian effects --- its initial stage merely assumes a small cosmological
constant without any high-density matter or radiation. If loop quantum gravity
could imply early-time stability only at the expense of requiring Planckian
physics, it would not be of much use in rescuing the no-boundary proposal.

Fortunately, several new properties conspire to solve these two problems in
one strike. In particular, it is possible to derive a consistent off-shell
$\beta$ as a function of the scale factor using only the
second-order equation.  (Technical details are provided in the supplementary
material.) 

Written as a function of $q(t)$, this new $\beta$ is given by
\begin{equation}
 \beta(t)=\frac{1-\delta^2M^{-2}
   \left(\ddot{q}+\frac{1}{2}\dot{q}^2/q\right)}{1-\frac{2}{3}
     \delta^2 \Lambda}\,,
\end{equation}
using small ${\rm d}g/{\rm d}q$ and ${\rm d}\kappa/{\rm d}q$. For
sub-Planckian $\Lambda$, the denominator is close to one. For small-$t$
no-boundary solutions, we have $q(t)\approx ct$ with constant $c>0$, and
\begin{equation} \label{beta}
 \beta\approx 1-\frac{\delta^2c}{2M^2} \frac{1}{t}<0
\end{equation}
is negative as long as $t<\frac{1}{2}\delta^2c/M^2$. Rather surprisingly, but
crucially, off-shell instantons in loop quantum cosmology imply signature
change even for sub-Planckian energy densities or $\Lambda$. The advantage of
the no-boundary proposal being independent of Planckian physics is therefore
not destroyed by bringing in signature change from loop quantum gravity.

{\em Timeless stability:} We are now in a position to draw our main
conclusions.  The results of \cite{NoSmooth,NoRescue} are based on a detailed
analysis of the Lorentzian path integral, paying special attention to suitable
contours when integrating over the lapse function. The path integral is
Lorentzian in that one integrates over real $N$, eliminating the Wick rotation
to imaginary $\tilde{N}=\pm iN$ in (\ref{ds}) suggested by the original
Hartle--Hawking proposal. Picard--Lefschetz theory is then used to improve the
convergence property of this highly oscillatory integral, shifting the
integration contour for $N$ into the complex plane (but not all the way to the
imaginary axis).

Since we are interested here in stability properties, we do not need an
explicit calculation of a path integral. However, we make use of a crucial
property observed and explained in \cite{NoRescue}: There are unbounded modes
$v$ of the metric which contribute to the path integral by an undamped
Gaussian exponential $\exp(\lambda v^2)$ with the ``wrong,'' positive sign of
the exponent, $\lambda>0$. Interpreting the path integral as a transition
amplitude, instability is implied because metric perturbations $v$ are not
suppressed. But how is it possible that the Lorentzian path integral,
integrating the bounded $\exp(iS/\hbar)$ for real lapse functions, results in
real undamped Gaussians for some modes? Picard--Lefschetz theory deforms the
integration contour such that certain complex $N$ are used in integrations,
but it does not change the value of the original real integral if there are no
poles in the complex plane.

A crucial insight of \cite{NoRescue} explains this puzzle by noting that the
integrand has a branch cut on the real $M$-axis. The contour must bypass this
branch cut by deviating slightly into the upper imaginary half-plane, and in
this way picks up complex actions from complex $M$. (The contour must be above
the branch cut to access the relevant saddle point which lies in the first
quadrant of $M$.)

Undamped Gaussians in the
Lorenztian path integral are a direct consequence of this branch cut.
Specifically, \cite{NoRescue} use the equation $\ddot{v}\approx
-\frac{1}{4}c^{-2}M^2\ell(\ell+2) v/t^2$ for a spherical harmonic tensor mode
$v$ of moment $\ell$, using a no-boundary background $q(t)=ct$. This equation
is solved by $v_{\pm}(t) = v_1t^{\frac{1}{2}(1\pm\gamma)}$ (with
$v_1=v_{\pm}(1)$) where $\gamma=\sqrt{1-\ell(\ell+2)M^2/c^2}$ clearly shows
the branch cut on the $M$-axis. Moreover, the action evaluated on the regular
solution $v_+$ is equal to $S_+(v_1)= \frac{1}{4}M^{-1} (\gamma- 1)v_1^2$ and
has a negative imaginary part above the branch cut.  Inserting the imaginary
$S_+$ in $\exp(iS/\hbar)$ leads to an undamped Gaussian in the path integral
of perturbations.  Our strategy is now to show that signature change from loop
quantum cosmology moves the branch cut to the imaginary $M$-axis, such that
the argument of \cite{NoRescue} no longer applies: The integrated action is
always real and does not lead to undamped Gaussians.

For small $t$, as shown in more detail in the supplementary material,
derivations of consistent perturbation equations in models of loop quantum
gravity \cite{ScalarHolInv} imply that we have the mode equation
\begin{equation} \label{vdd}
 \ddot{v}\approx \frac{1}{4}\left((n-2\epsilon)(n+2)+\epsilon(\epsilon+2) -
   \beta \frac{M^2\ell(\ell+2)}{c^2}\right) \frac{v}{t^2}
\end{equation}
with $\epsilon=-1$ for our $\beta$ while $\epsilon=0$ (and $n=0$) classically.
We still have solutions $v_{\pm} = v_1t^{\frac{1}{2}(1\pm\gamma)}$, but now
\begin{equation} \label{gamma}
 \gamma=\sqrt{1+n(n+2)-\beta\frac{\ell(\ell+2)M^2}{c^2}}\,.
\end{equation}
With dynamical signature change in an intermediate range of $t$, that is
$\beta<0$, $\gamma$ is always real for real $M$. Its branch cuts in the
complex plane are now on the imaginary $M$-axis where they do not affect the
Lorentzian path integral, while the qualitative structure of saddle points
remains unchanged. The argument for Gaussians with positive exponents, given
in \cite{NoRescue}, no longer applies, and indeed the action $S_+$ is always
real and finite.

Most importantly, in the asymptotic regime of small $t$ using (\ref{beta}), the
dominant term in (\ref{vdd}) is $\ddot{v}\approx \frac{1}{4}\alpha v/t^3$ with
$\alpha=2\delta^2\ell(\ell+2)/c>0$. This equation has a regular solution given
by a modified Bessel function of the second kind, $v(t)=v_1 \sqrt{t}
K_1(\sqrt{\alpha/t}) /K_1(\sqrt{\alpha})$. The action, 
\begin{equation}
 S(v_1)=\frac{\sqrt{\alpha}}{4M}
\frac{K_0(\sqrt{\alpha})}{K_1(\sqrt{\alpha})} v_1^2\,,
\end{equation}
is again real and finite and no undamped Gaussians result.  Signature change
is crucial here because these properties would be different if $\alpha<0$, as
is the case for $\beta>0$.

Our new combination of the Lorentzian path integral with results from loop
quantum cosmology is surprisingly productive. Loop quantum cosmology has led
to an unexpected new property of quantum space-time by suggesting
non-singular, dynamical signature change in the early universe. This effect
strengthens the original intuition behind the no-boundary proposal by
replacing the technical notion of Euclidean path integrals with dynamical
signature change embodied by an effective line element (\ref{dsSig}).
Moreover, it is just what the no-boundary proposal, or any theory of a smooth
beginning, needs in order to imply stable perturbations without imposing final
conditions. While these are quantum-gravity effects, they do not require a
Planckian energy density or cosmological constant.

\noindent {\em Acknowledgements:}
The authors are grateful to Job Feldbrugge, Jean-Luc Lehners, and Neil Turok
for discussions, and to the Perimeter Institute for Theoretical Physics for
hospitality. Research at Perimeter Institute is supported by the Government of
Canada through Innovation, Science and Economic Development, Canada and by the
Province of Ontario through the Ministry of Research, Innovation and Science.
This work was supported in part by NSF grant PHY-1607414. S.B. is supported in
part by the Ministry of Science, ICT \& Future Planning, Gyeongsangbuk-do and 
Pohang City and the National Research Foundation of Korea grant no. 2018R1D1A1B07049126.

\section{Supplementary material}

{\em Sample solution of background equation:} If one ignores holonomy
modifications ($\delta=0$), an analytical solution of
\begin{equation} \label{qdd1}
\ddot{q}=\frac{2}{3}\Lambda M^2 \left(1-\frac{3}{\Lambda} \frac{{\rm
		d}\kappa}{{\rm d}q}
+\delta^2\left(1-\frac{3}{\Lambda}\frac{\kappa}{q}\right) \left(2\frac{{\rm
		d}\kappa}{{\rm d}q}-\frac{\kappa}{q}-\frac{\Lambda}{3}\right)\right)
\,.
\end{equation}
for $q(t)$ with no-boundary conditions $q(0)=0$ and $q(1)=q_1$ can be found if
$\kappa(q)=\kappa_0 q^2$ is quadratic. Its small-$t$ behavior
\begin{equation}\label{qSoln}
q(t)\approx \frac{2M\sqrt{\kappa_0}}{\sin(2\sqrt{\kappa_0}M)}
\left(q_1+ \frac{\Lambda}{6}\left(\cos(2\sqrt{\kappa_0}M)-1\right)\right)t 
\end{equation}
is consistent with the general form used in the main text.

{\em Perturbations:} Covariant equations compatible with the modified
Friedmann equation used in the main text have been shown to have very similar
features in a variety of models \cite{DeformedCosmo}. The explicit example we
have used has been derived in \cite{ScalarHolInv} for cosmological
perturbations around spatially flat isotropic models.  These results assume
that $g(q)=g_0 q^n$ is an integer power law for small $q$, and imply that
tensor perturbations $h(\eta,x)$ in conformal time $\eta$ are subject to
\begin{equation} \label{h}
h''+\left(2(1+n)\frac{a'}{a}-\frac{\beta'}{\beta}\right) h'-
\frac{\beta}{2\Sigma} (1+n) \nabla^2h=0\,,
\end{equation}
where $\beta$ is the same function that appears in the effective line element,
and $\Sigma=\frac{1}{2}g_0^{-1} (3-2n)(3-n)$ in terms of the parameters
already introduced.  We should transform this equation to the variable $q=a^2$
instead of $a$, and a time coordinate that corresponds to the choice
$N=M/a$. Substitution shows that (\ref{h}) is equivalent to
\begin{equation} \label{v}
\ddot{v}-\frac{\ddot{z}}{z} v-\frac{M^2}{q^2}\beta \nabla^2v=0
\end{equation}
where $z=q^{1+n/2}/\sqrt{|\beta|}$ and $v=zh$. For small $t$ and a no-boundary
off-shell instanton such that $q(t)\sim ct$, we 
have the mode equation
\begin{equation} \label{vdd1}
\ddot{v}\approx \frac{1}{4}\left((n-2\epsilon)(n+2)+\epsilon(\epsilon+2) -
\beta \frac{M^2\ell(\ell+2)}{c^2}\right) \frac{v}{t^2}
\end{equation}
for a spherical harmonic of moment $\ell$, with $\epsilon=-1$ if there is
signature change, while $\epsilon=0$ classically.

{\em Second-order equation:} Canonical derivations show that the
signature-change function $\beta$, depending on the background momentum $P$,
has the general form $\beta(P)=\cos(2\delta P)$
\cite{JR,ScalarHolInv,DeformedCosmo}. The modified constraint generates
coupled first-order equations for $Q$ and $P$, in which trigonometric
identities can be used to obtain $\cos(2\delta P)$ as a function of $Q$ and
its time derivative. This is the usual way of deriving $\beta(P)$ in loop
quantum cosmology.

There are two technical problems if one tries to develop a new derivation for
off-shell instantons. First, one should not use the first-order equations but
only the second-order equation which is also obtained from the modified
constraint but has a larger solution space. In particular, it contains
no-boundary solutions which do not solve the first-order equations. In fact,
if we formally tried to derive $\beta=\cos(2\delta P)$ from the first-order
equations for a no-boundary solution, we would not obtain a real value.

It turns out that the second-order equation that follows from the modified
constraint is simpler than could have been expected, in that
\begin{widetext}
	\begin{equation} \label{Qdd}
	\ddot{Q}
	= \frac{3}{2}\frac{M^2}{\delta^2} \frac{Q^{1/3}}{(4\pi G)^{2/3}} \left(1-\left(1
	+ \delta^2Q^{1/3} 
	\left(\frac{3}{(4\pi G)^{2/3}}\frac{{\rm d}\kappa}{{\rm d}Q}- 
	2 \frac{{\rm d}(mgQ^{-1/3})}{{\rm d}Q}\right)\right) \beta
	\right)
	\end{equation}
\end{widetext}
is linear in $\beta$. No square root is required when solving for $\beta$ in
terms of $Q$ and its time derivatives, and therefore $\beta$ is always
real. However, the solution is not guaranteed to be such that $|\beta|\leq 1$,
as one would expect if $\beta=\cos(2\delta P)$. This property is not
problematic because, having forgone the first-order formulation, we do not
need a real $P$ for which  $|\beta|\leq 1$. These results guarantee that
signature change is consistent also for off-shell instantons.
In the regime of interest, we obtain
\begin{equation}
\beta=\frac{1-\delta^2M^{-2}
	\left(\ddot{q}+\frac{1}{2}\dot{q}^2/q\right)}{1-\frac{2}{3}
	\delta^2 \Lambda}
\end{equation}
from (\ref{Qdd}), as used and analyzed in the main text.

{\em Effect of inverse-$a$ corrections:} We can use the background solution
(\ref{qSoln}) to demonstrate the role played by inverse-$a$ corrections in the
rescue maneuver of signature change.  Using (\ref{vdd1}) with inverse-$a$
corrections, we note that they, too, improve stability even without signature
change. However, they imply a real $\gamma$ only for low multipole moments,
for small values of $\ell$ depending on the ambiguity parameter $n$. 

The background solution (\ref{qSoln}) indicates another consequence of
inverse-$a$ corrections: If we insert the specific value of $c$ implied by
this solution, $M^2$ in (\ref{vdd1}) cancels out and is replaced by a bounded
function of $M$. This result further stabilizes low-$\ell$ modes because the
branch cuts in the $M$-plane may not be reached for such values. In the
absence of additional closed-form background solutions, however, it is
difficult to generalize this latter argument to general power laws of
inverse-$a$ corrections. In any case, we learn that, if there is signature
change, including inverse-$a$ corrections does not weaken our conclusion of
stable perturbations.

\end{document}